\documentclass{ws-ijmpe}

\begin{document}

\markboth{William A. Horowitz}{LHC Predictions from an extended theory with Elastic, Inelastic, and Path Length Fluctuating Energy Loss}

\catchline{}{}{}{}{}
\newcommand{\be}{\begin{equation}}
\newcommand{\ee}{\end{equation}}
\newcommand{\bfig}{\begin{figure}}
\newcommand{\efig}{\end{figure}}
\newcommand{\bea}{\begin{eqnarray}}
\newcommand{\eea}{\end{eqnarray}}
\newcommand{\infinitessimal}{\mathrm{d}}
\newcommand{\infinitesmal}{\mathrm{d}}
\newcommand{\infinitesimal}{\mathrm{d}}
\newcommand{\intd}{\mathrm{d}}


\newcommand{\raa}{$R_{AA}$ }
\newcommand{\raacomma}{$R_{AA}$}
\newcommand{\raaphi}{$R_{AA}(\phi)$ }
\newcommand{\raaphicomma}{$R_{AA}(\phi)$} 
\newcommand{\raapt}{$R_{AA}(\eqnpt)$ }
\newcommand{\raaptcomma}{$R_{AA}(\eqnpt)$} 
\newcommand{\raaphipt}{$R_{AA}(\phi;\,\eqnpt)$ }
\newcommand{\raaphiptcomma}{$R_{AA}(\phi;\,\eqnpt)$} 
\newcommand{\RAA}{\raa}
\newcommand{\RAAcomma}{\raacomma}
\newcommand{\RAAphi}{\raaphi}
\newcommand{\RAAphicomma}{\raaphicomma}
\newcommand{\RAAphipt}{\raaphipt}
\newcommand{\RAAphiptcomma}{\raaphiptcomma}
\newcommand{\vtwo}{$v_2$ }
\newcommand{\vtwocomma}{$v_2$}
\newcommand{\eqnraa}{R_{AA}}
\newcommand{\eqnraaphi}{R_{AA}(\phi)}
\newcommand{\eqnraaphipt}{R_{AA}(\phi;\,\eqnpt)} 
\newcommand{\eqnRAA}{\eqnraa} 
\newcommand{\eqnraapt}{R_{AA}(\eqnpt)}

\newcommand{\eqnRAAphi}{\eqnraaphi}
\newcommand{\eqnRAAphipt}{\eqnraaphipt}
\newcommand{\eqnvtwo}{v_2}
\newcommand{\vtwovsraa}{\vtwo vs.~\raa}
\newcommand{\vtwovsraacomma}{\vtwo vs.~\raacomma}

\newcommand{\pp}{$p+p$ }
\newcommand{\ppcomma}{$p+p$}
\newcommand{\dau}{$d+Au$ }
\newcommand{\daucomma}{$d+Au$}
\newcommand{\auau}{$Au+Au$ }
\newcommand{\auaucomma}{$Au+Au$}
\newcommand{\aplusa}{$A+A$ }
\newcommand{\aplusacomma}{$A+A$}
\newcommand{\cucu}{$Cu+Cu$ }
\newcommand{\cucucomma}{$Cu+Cu$}

\newcommand{\rhopart}{$\rho_{\textrm{\footnotesize{part}}}$ }
\newcommand{\rhopartcomma}{$\rho_{\textrm{\footnotesize{part}}}$}
\newcommand{\eqnrhopart}{\rho_{\textrm{\footnotesize{part}}}}
\newcommand{\npart}{$N_{\textrm{\footnotesize{part}}}$ }
\newcommand{\npartcomma}{$N__{\textrm{\footnotesize{part}}}$}
\newcommand{\eqnnpart}{N_{\textrm{\footnotesize{part}}}}
\newcommand{\taa}{$T_{AA}$ }
\newcommand{\taacomma}{$T_{AA}$}
\newcommand{\eqntaa}{T_{AA}}
\newcommand{\rhocoll}{$\rho_{\textrm{\footnotesize{coll}}}$ }
\newcommand{\rhocollcomma}{$\rho_{\textrm{\footnotesize{coll}}}$}
\newcommand{\eqnrhocoll}{\rho_{\textrm{\footnotesize{coll}}}}
\newcommand{\ncoll}{$N_{\textrm{\footnotesize{coll}}}$ }
\newcommand{\ncollcomma}{$N_{\textrm{\footnotesize{coll}}}$}
\newcommand{\eqnncoll}{N_{\textrm{\footnotesize{coll}}}}
\newcommand{\dndy}{$\frac{dN_g}{dy}$ }
\newcommand{\dndycomma}{$\frac{dN_g}{dy}$}
\newcommand{\eqndndy}{\frac{dN_g}{dy}}
\newcommand{\eqndndyabs}{\frac{dN_g^{abs}}{dy}}
\newcommand{\eqndndyrad}{\frac{dN_g^{rad}}{dy}}
\newcommand{\dnslashdy}{$dN_g/dy$ }
\newcommand{\dnslashdycomma}{$dN_g/dy$}
\newcommand{\eqndnslashdy}{dN_g/dy}
\newcommand{\as}{\alpha_s}
\newcommand{\alphas}{$\as$ }
\newcommand{\alphascomma}{$\as$}
\newcommand{\eqnalphas}{\as}

\newcommand{\pt}{$p_\perp$ }
\newcommand{\pT}{\pt}
\newcommand{\ptcomma}{$p_\perp$}
\newcommand{\pTcomma}{\ptcomma}
\newcommand{\eqnpt}{p_\perp}
\newcommand{\lowpt}{low-\pt}
\newcommand{\lowptcomma}{low-\ptcomma}
\newcommand{\midpt}{intermediate-\pt}
\newcommand{\midptcomma}{intermediate-\ptcomma}
\newcommand{\highpt}{high-\pt}
\newcommand{\highptcomma}{high-\ptcomma}
\newcommand{\Aperp}{$A_\perp$ }
\newcommand{\Aperpcomma}{$A_\perp$}
\newcommand{\eqnAperp}{A_\perp}
\newcommand{\rperp}{$r_\perp$ }
\newcommand{\rperpcomma}{$r_\perp$}
\newcommand{\eqnrperp}{r_\perp}
\newcommand{\eqnrperpHS}{r_{\perp,HS}}
\newcommand{\eqnrperpWS}{r_{\perp,WS}}
\newcommand{\Rperp}{$R_\perp$ }
\newcommand{\Rperpcomma}{$R_\perp$}
\newcommand{\eqnRperp}{R_\perp}

\newcommand{\pizero}{$\pi^0$ }
\newcommand{\eqnpizero}{\pi^0}

\newcommand{\qhat}{$\hat{q}$ }
\newcommand{\qhatcomma}{$\hat{q}$}
\newcommand{\eqnqhat}{\hat{q}}

\newcommand{\infinity}{\infty}

\newcommand{\eq}[1]{Eq.~(\ref{#1})}
\newcommand{\fig}[1]{Fig.~\ref{#1}}
\newcommand{\tab}[1]{Table \ref{#1}}
\newcommand{\captionsize}{\small}
\title{LHC Predictions from an extended theory with Elastic, Inelastic, and Path Length Fluctuating Energy Loss}

\author{\footnotesize William A. Horowitz}

\address{Department of Physics, Columbia University, 538 W. 120$^{th}$ St.\\
New York, NY 10027, USA\\
Frankfurt Institute for Advanced Studies (FIAS), Johann Wolfgang Goethe University, Max-von-Laue-Str. 1\\
60438 Frankfurt am Main, Germany\\
horowitz@phys.columbia.edu}

\maketitle

\begin{history}
\received{(received date)}
\revised{(revised date)}
\end{history}

\begin{abstract}
We present the LHC predictions for the WHDG model of radiative, elastic, and path length fluctuating energy loss.  We find the \pt dependence of \raa is qualitatively very different from AWS-based energy loss extrapolations to the LHC; the large \pt reach of the year one data at the LHC should suffice to distinguish between the two.  We also discuss the importance of requiring a first elastic scatter before \emph{any} medium-induced elastic or radiative loss occurs, a necessary physical effect not considered in any previous models.
\end{abstract}

\section{Introduction}
Currently, several papers\cite{Wicks:2005gt,Eskola:2004cr,Dainese:2004te,Vitev:2005he,Wang:2004yv,Turbide:2005fk} claim that their respective descriptions of energy loss properly account for the high-\pt suppression of central pion data observed at RHIC\cite{Isobe:2005mh,Shimomura:2005en,Adler:2003qi}.  Other RHIC observables have been suggested as a means of differentiating between these, such as their predictions for the centrality dependence, back-to-back jet quenching, two and three particle correlations, photon spectra, etc.  We suggest that the large \pt lever arm of pion data at the LHC will distinguish between the WHDG and the AWS models.  We will not discuss \cite{Wang:2004yv,Turbide:2005fk} further; higher twist predictions for the LHC do not seem to have been calculated, and AMY LHC results go out to only low \pt ($\sim 30$ GeV).

The important quantity in any of these calculations is $P(\epsilon)$, the probability that the final momentum is some fraction of the initial momentum, $p_{\perp,f}=(1-\epsilon)p_{\perp,i}$.  For calculations that include only radiative processes, $P(\epsilon)=P_{rad}(\epsilon)$, where
\be
\label{prad}
P_{rad}(x)=\left\{\begin{array}{l}
P_0^g\delta(x)+\widetilde{P}_{rad}(x)\\
e^{-N_g}\sum\limits_n\frac{N_g^n}{n!}\widetilde{P}_{rad,n}(x) = 
e^{-N_g}\delta(x)+ e^{-N_g} N_g \widetilde{P}_{rad,1}(x)+\ldots.
\end{array}\right.
\ee
The possibility that no gluons are emitted (subsequent to the original gluon radiation created by the initial hard scatter) is encapsulated in the coefficient, $P_0^g$, of the $\delta(x)$ term.  

Mustafa showed that the elastic contribution to charm energy loss is comparable to the radiative loss\cite{Mustafa:2004dr}.  This work was extended\cite{Wicks:2005gt} to show that, despite long held assumptions to the contrary, elastic energy loss is in fact not negligible for all parton jet flavors.  The WHDG model includes elastic energy losses in its \raa calculations by convolving the two probability distributions,
\be
P(\epsilon)=\int \! dx P_{rad}(x)P_{el}(\epsilon-x).
\ee

However, none of these previous models include the possibility that the parton jet escapes the medium completely unmodified, $P_0=\exp(-N_c)$, where $N_c$ is the average number of elastic collisions suffered along the parton path.  Note that this is different from, and in addition to, the probability of radiating no gluons, $P_0^g=\exp(-N_g)$, included in the $P_{rad}$ term in \eq{prad}.  

Due to the approximations used in the radiative calculations overabsorption, $P(\epsilon>1)$, has a large support for highly suppressed jets.  Usually, one of two prescriptions is applied to remove this unphysical artifact.  Either the integrated excess probability weighs an explicit delta function,
\be
\label{norw}
P(\epsilon)=P_{old}(\epsilon)\theta(1-\epsilon) + \int_1^\infinity \!dxP_{old}(x) \delta(1-\epsilon),
\ee
or reweighs (rw) the original distribution,
\be
\label{rw}
P_{rw}(\epsilon)=\frac{1}{\int_1^\infinity \!dxP_{old}(x)}P_{old}(\epsilon)\theta(1-\epsilon).
\ee
Clearly the latter approach leads to larger \raa values for the two.  For large overabsorption energy loss details are lost to the removal of the unphysical $\epsilon>1$ region.  
\section{WHDG Model}
In our extended theory of elastic loss in addition to radiative loss calculated in a realistic geometry, we take
\be
\eqnraa(\eqnpt)\approx \langle\int d\epsilon (1-\epsilon)^{n(\eqnpt)}P(\epsilon)\rangle_{geom}.
\ee
$n(\eqnpt)$ comes from the power law approximation to the pQCD production spectrum (minus either one or two depending on the Jacobian).   $P(\epsilon)$ is the convolved probability distribution $P(\epsilon|\vec{x},\phi,\eqnpt)=\int dx P_{rad}(x|\vec{x},\phi,\eqnpt)P_{el}(\epsilon-x|\vec{x},\phi,\eqnpt)$.  And $\langle\ldots\rangle_{geom}$ corresponds to geometrical averaging, 
\be
\label{lfixed}
\langle \ldots \rangle_{geom} = 
\left\{ \begin{array}{l} 
\int dL\ldots \delta(L-\overline{L}) \\
\int dL\ldots P(L); \end{array} \right.
\ee
see \cite{Wicks:2005gt} and references therein for details.  

\begin{figure}[h]
\vspace{.15in}
\begin{center}
$\begin{array}{c@{\hspace{.15in}}c}
\epsfxsize=.5\columnwidth
\epsffile{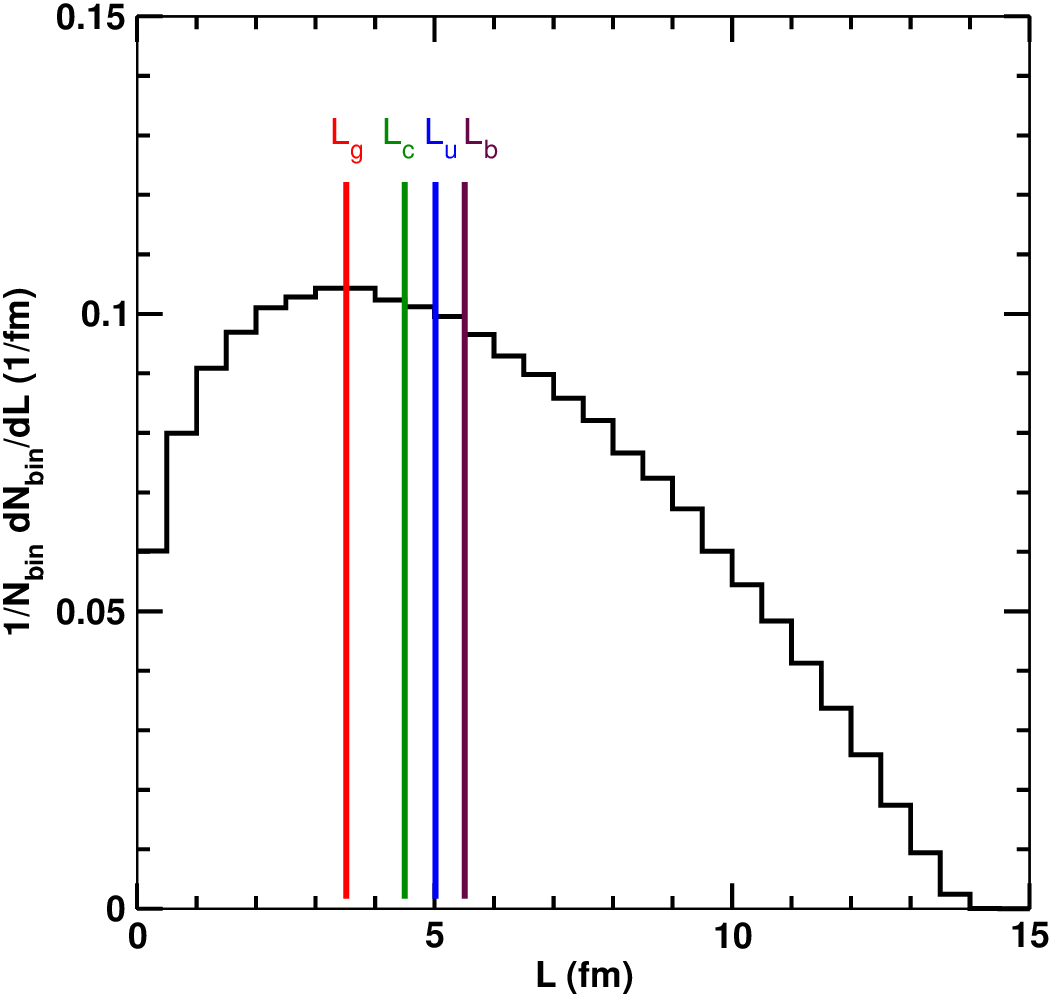 } & \epsfxsize=.5\columnwidth
\epsffile{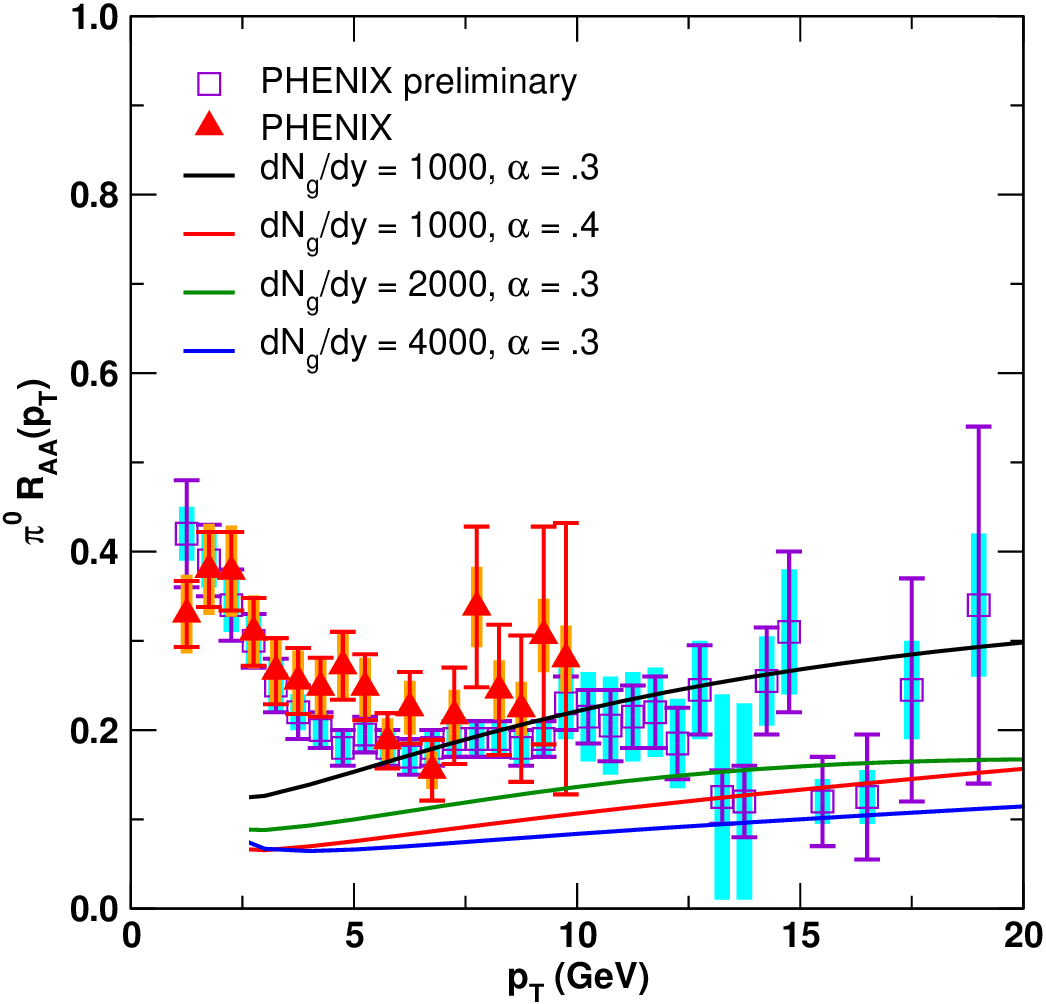 } \\ [-.11in]
{\mbox {\tiny {\bf (a)}}} & {\mbox {\tiny {\bf (b)}}}
\end{array}$
\end{center}
\vspace{-.1in}
\caption{All plots depict midrapidity for most-central collisions. (a) The histogram gives the distribution of path lengths traversed by hard scatterers.  The lengths, $L(\vec{x}_\perp,\phi)$, are weighted by the probability of production and averaged over azimuth; these quantities were found using the Glauber thickness function from a realistic Woods-Saxon base geometry.  A purely geometric quantity, $P(L)$ is the same for all jet varieties.  However, no single $\overline{L}_{fit}$ can best reproduce the results of the full distribution for all parton jet flavors. (b) The WHDG model reproduces the PHENIX pion \raapt data$^{7,8,9}$ for the conservative, entropy-constrained input parameters of $dN_g/dy=1000$ and $\eqnalphas=.3$.  Model sensitivity to varying the medium density is much greater than that observed in $^2$; however, the large sensitivity to changes in \alphas suggests a need for relaxing the fixed coupling constant approximation in future work.}
\label{fig1}
\end{figure}

The first line of \eq{lfixed} is a simplification of full geometrical averaging, and presents one with a number of choices for $\overline{L}$.  The three most natural possibilities are
\be
\overline{L}=\left\{
\begin{array}{l}
\overline{L}_{prod}\\
\overline{L}_{obs}(\eqnpt)\\
\overline{L}_{fit},
\end{array}\right.
\ee
where $\overline{L}_{prod}$ is the average length that all produced hard partons travel, $\overline{L}_{prod}=\frac{1}{2\pi}\int\!d^2\vec{x}d\phi \rho_{prod} L(\vec{x},\phi)$, $\overline{L}_{obs}(\eqnpt)$ is the average length that an observed jet travelled on its way out of the medium, $\overline{L}_{obs}(\eqnpt)=\frac{1}{2\pi}\int\!d^2\vec{x}d\phi \rho_{prod} L(\vec{x},\phi) \int\!d\epsilon (1-\epsilon)^n(\eqnpt)P(\epsilon|\vec{x},\phi,\eqnpt)$, and $\overline{L}_{fit}$ is the single fixed length that best approximates the full geometry average, given by the second line of \eq{lfixed}.  While it is generally impractical to use $\overline{L}_{fit}$ in calculations (one must first compute the proper geometry average first anyway in order to find $\overline{L}_{fit}$), one sees from \fig{fig1} that employing a single length without calculating the full geometry average is problematic; there is no \emph{a priori} method to estimate the degree to which using a single, representative length deviates from the full solution.  In fact, \fig{fig1} shows the $\overline{L}_{fit}$ hierarchy differs from one's naive expectations.
\input{epsf}
\section{RHIC Results}                 
We see from \fig{fig1} that for the conservative \alphas of $.3$ and entropy-constrained $\eqndnslashdy=1000$, the WHDG model is consistent with the RHIC pion data observed by PHENIX\cite{Isobe:2005mh,Shimomura:2005en,Adler:2003qi}.  While \cite{Horowitz:2006ya} saw these results as sensitive enough to changes in the input to be considered not fragile (unlike the conclusions of \cite{Eskola:2004cr}), a more sophisticated statistical analysis showed that, when requiring the rather high 90\% confidence level, the combination of theoretical sensitivity and experimental error results in an approximately factor of 2 range of allowable medium densities\cite{Sahlmueller:2007wx}.  We note that previous, radiative only GLV-based energy loss models also are consistent with the data\cite{Vitev:2005he}; in comparison, the addition of elastic losses does not oversuppress our results for two reasons: we also include geometrical path length fluctuations, and we use a fixed \alphascomma.  \cite{Vitev:2005he} incorporates some running of the coupling.  Due to the low $q_\perp$ and $k_\perp$'s involved, a running \alphas will often be evaluated at its cutoff; for \cite{Vitev:2005he}, $\alpha_s^{max}=.5$.  We note that the inclusion of elastic loss allows for a small \raa without much overabsorption in $P_{rad}$, $P_{el}$, or their convolution.

In order to understand the qualitative differences that will arise in the LHC predictions, we briefly discuss the RHIC results from AWS-type models\cite{Eskola:2004cr,Dainese:2004te}.  To fit the pion suppression, a nonperturbatively large \qhat is used.  Specifically, \qhat is proportional to the three-fourth's root of the energy density of the medium, $\eqnqhat=c\epsilon^{3/4}$, where pQCD estimates\cite{Baier:2002tc} give $c\sim2$; rather, $c\sim8-20$ is needed in \cite{Eskola:2004cr,Dainese:2004te}.  This is due to the combination of radiative only energy loss, oversimplified treatment of geometry\cite{Horowitz:2006ya}, and the unitarity-violating $P_0^g>1$ in their model\cite{Vitev:2005he}.  
\section{LHC Predictions}
We show in \fig{lhcraa} the large qualitative difference in \pt dependence for predicted LHC pion \raa from the WHDG model and from two different implementations of the AWS model; one easily sees the dramatic rise in \raa with increasing \pt from the first as opposed to the flat in \pt results of the latter two.  The consistency of Vitev's curve\cite{Vitev:2005he} with ours over a range of $dN_g/dy$ and the consistency of the two AWS calculations\cite{Eskola:2004cr,Dainese:2004te} suggest that this is a robust result.  The origin of the difference can be easily understood given our discussion of RHIC results.  For the case of WHDG, the pion \raa at RHIC does not require much overabsorption.  The modest increase in medium density, $\sim2-3$ based on either an extrapolation of PHOBOS results\cite{Back:2001ae,Adcox:2000sp} or predictions from the CGC\cite{Kharzeev:2004if,McLerran:2004fg}, for the LHC leads to small energy losses at high momenta that can be well approximated by the pocket asymptotic energy loss formulae
\bea
\epsilon_{rad}=\Delta E_{rad}/E & \sim & \eqnalphas^3\log(E/\mu^2L)/E\\
\epsilon_{el}=\Delta E_{el}/E & \sim & \eqnalphas^2\log(\sqrt{ET}/m_g)/E.
\eea
As \pt increases the $\log(E)/E$ reduction in energy loss is not compensated by the slow (in comparison to RHIC) increase in the power law, $n(\eqnpt)$, partonic production spectrum; thus \raa increases with \ptcomma.  On the other hand, the AWS mimics the small normalization of the RHIC data by highly suppressing their jets; the details of energy loss are lost (in the delta functions at zero and one, or at zero and the reweighting of $P$), thus flattening the results.  Moreover, the two AWS models represented in \fig{lhcraa} used EKRT-type medium density scaling\cite{Eskola:1999fc}; this makes the LHC $\sim7$ times more dense than RHIC.  The LHC jets are thus even more dramatically overabsorbed and, again, flat in \ptcomma.  
\bfig[!h]
\begin{center}
\leavevmode
\includegraphics[width=.5 \columnwidth]{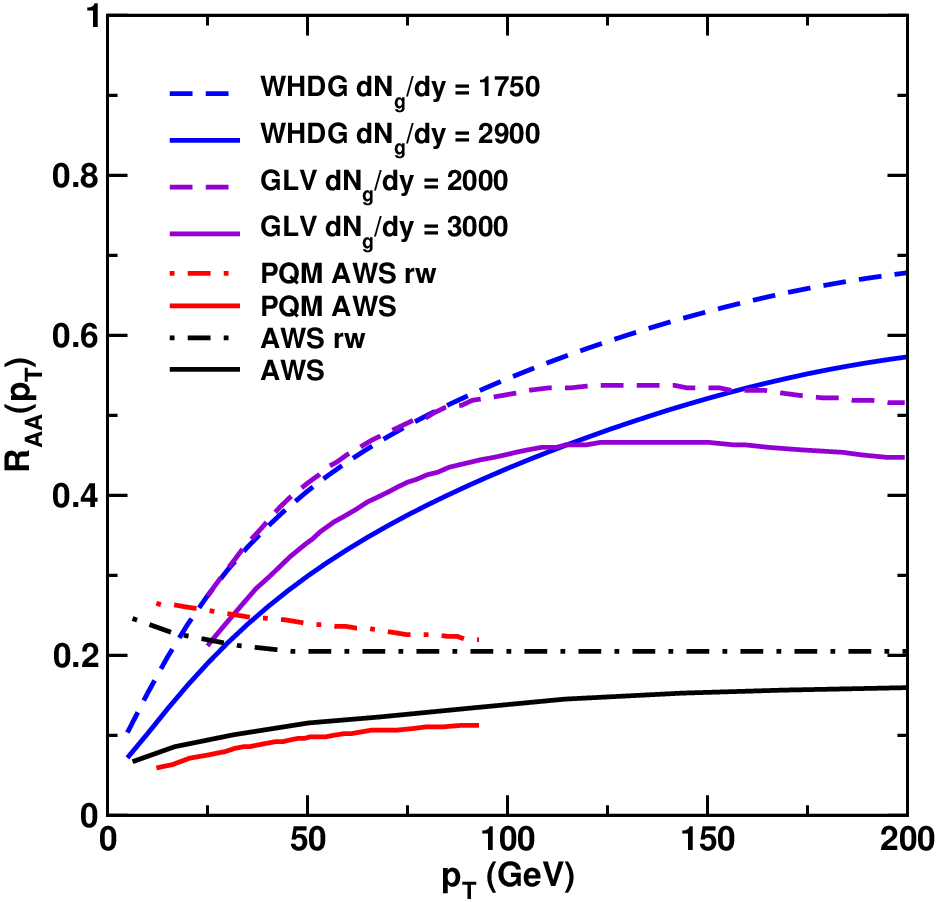}
\end{center}
\vspace{-.2in}
\caption{LHC predictions for several energy loss models.  WHDG curves correspond to $^1$, GLV to $^4$, PQM AWS to $^3$, and AWS to $^2$.  For the latter two, rw indicates the use of \eq{rw} for overabsorption; otherwise \eq{norw} was employed.  Also for the latter two, $\eqnqhat = 100$ and $\eqnqhat = 68$ were the values of the AWS input parameter, respectively.  Notice the sharp rise in \pt for the WHDG and GLV curves as opposed to the flatness of the AWS and PQM AWS results.}
\label{lhcraa}
\efig 
\vspace{-.3in}
\section{Probability of Nothing}
An outgoing parton jet must encounter \emph{at least} one scatter in order for medium-induced energy loss to occur.  There exists, then, a new (in the sense that it has not been considered before) probability of no in-medium energy loss, $P_0=\exp(-N_c)$, where $N_c=\int\!dz\sigma_{el}(z)\rho(z)$ is the average number of elastic collisions for the jet.  Note that $N_c$ has a strong dependence on the Casimir associated with the quark or gluon jet.  After this first elastic collision the total probability of energy loss is then the convolution of the elastic and radiative processes:
\be
P(\epsilon)=P_0\delta(\epsilon)+(1-P_0)\int\!dxP_{rad}(x)P_{el}(\epsilon-x).
\ee
In the above equation, this $P_0\delta(\epsilon)$ piece is in addition to the probability of no glue emission, $P_0^g\delta(x)$, in $P_{rad}(x)$, \eq{prad}.  

For fixed $\eqnalphas=.3$, including $P_0$ physics accounts for 50\% of \raacomma.  Allowing $\eqnalphas(T)$ to run as $\eqnalphas (q^2 = 2 \pi T(z))$ reduces $P_0$ by a factor of 2.  Finally, integration over momentum transfers with $\eqnalphas(q^2)$ given by vacuum running formally gives $P_0=0$.
\section{Conclusions}
We see from \fig{lhcraa} that the LHC $\eqnraa(\eqnpt)$ pion data will distinguish between energy loss models.  While WHDG, with its radiative, elastic, and path length fluctuations, predicts a significant rise as a function of \ptcomma, AWS-type models predict flat \pt dependence.  This flatness in \pt of the latter is due to the high suppression of their jets, which forces the energy loss details into the region of overabsorption.  Contrarily, the former predicts a much smaller fractional energy loss for jets; the high-\pt behavior is well approximated by the analytic asymptotic energy loss formulae and is thus responsive to their details.  We also found that moderate in opacity \raa predictions are sensitive to noninteracting free jets, whose influence has not been considered previously.

\end{document}